\documentclass[12pt]{article}
\begin {document}

\begin{center}
{\bf COMPARISON OF $k_T$-FACTORIZATION APPROACH AND QCD
PARTON MODEL FOR CHARM AND BEAUTY HADROPRODUCTION}

\vspace{0.5cm}

M.G.Ryskin and A.G.Shuvaev \\
Petersburg Nuclear Physics Institute, \\
Gatchina, St.Petersburg 188350 Russia \\

\vspace{0.3cm}

Yu.M.Shabelski \\
The Abdus Salam International Centre \\
for Theoretical Physics, Trieste, Italy \\
and \\
Petersburg Nuclear Physics Institute, \\
Gatchina, St.Petersburg 188350 Russia\footnote{Permanent
address}
\end{center}

\vspace{0.5cm}

\begin{abstract}
We compare the numerical predictions for heavy quark production in high
energy hadron collisions of the conventional QCD parton model and the
$k_T$-factorization approach (semihard theory). The total production
cross sections, one-particle rapidity and $p_T$ distributions as well as
two-particle correlations are considered. The difference in the
predictions of the two approaches is not very large, while the shapes of
the distributions are slightly different.

\end{abstract}

\vspace{2cm}

E-mail RYSKIN@THD.PNPI.SPB.RU

E-mail SHABELSK@THD.PNPI.SPB.RU

E-mail SHUVAEV@THD.PNPI.SPB.RU

\newpage

\section{Introduction}

The investigation of heavy quark production in high energy hadron
collisions is an important method for studying the quark-gluon
structure of hadrons. Realistic estimates of the cross section of
heavy quark production as well as their correlations are necessary in
order to plan experiments on existing and future accelerators as
well as in cosmic ray physics.

The description of hard interactions in hadron collisions within the
framework of QCD is possible only with the help of some phenomenology,
which reduces the hadron-hadron interaction to the parton-parton one via
the formalism of the hadron structure functions. The cross sections of
hard processes in hadron-hadron interactions can be written as the
convolutions squared matrix elements of the sub-process calculated
within the framework of QCD, with the parton distributions in the
colliding hadrons.

The most popular and technically simplest approach is the so-called QCD
collinear approximation, or parton model (PM). In this model all
particles involved are assumed to be on mass shell, carrying only
longitudinal momenta, and the cross section is
averaged over two transverse polarizations of the incident gluons.
The virtualities $q^2$ of the initial partons are taken into account
only through their structure functions. The cross sections of QCD
subprocess are calculated usually in the leading order (LO), as well
as in the next to leading order (NLO) \cite{1,2,NDE,Beer,Beer1}. The
transverse momenta of the incident partons are neglected in the QCD
matrix elements. This is the direct analogy of the Weizsaecker-Williams
approximation in QED. It allows to describe quite reasonably the
experimental data on the total cross sections and one-particle
distributions of produced heavy flavours, however it can not reproduce,
say, the azimuthal correlations \cite{MNR} of two heavy quarks, as well
as the distributions of the total transverse momentum of heavy quarks
pairs \cite{FMNR}, which are determined by the transverse momenta
of the incident partons.

There is an attempt to incorporate the transverse momenta of the incident
partons by a random shift of these momenta ($k_T$ kick) \cite{FMNR}
according to certain exponential distributions. This allows to describe
quantitatively the two-particle correlations \cite{FMNR}, but it
creates the problems in the simultaneous description of one-particle
longitudinal and transverse momentum distributions \cite{Shab}.

Another possibility to incorporate the incident parton transverse
momenta is referred to as $k_T$-factorization approach \cite{CCH,CE,MW,CH,CC},
or the theory of semihard interactions \cite{GLR,LR,8,lrs,3,SS}. Here the
Feynman diagrams are calculated taking account of the virtualities and of
all possible polarizations of the incident partons. In the small $x$
domain there are no grounds to neglect the transverse momenta of the
gluons, $q_{1T}$ and $q_{2T}$, in comparison with the quark mass and
transverse momenta, $p_{iT}$. Moreover, at very high energies and
very high $p_{iT}$ the main contribution to the cross sections comes
from the region of $q_{1T} \sim p_{1T}$ or $q_{2T} \sim p_{1T}$.
The QCD matrix elements of the sub-processes are rather complicated in
such an approach. We have calculated them in the LO. On the other hand,
the multiple emission of soft gluons is included here. That is why
the question arises as to which approach is more constructive.

Most of the published papers on $k_T$-factorization have presented no
numerical results or presented rather incomplete ones.
Old sets of structure functions have been used, and, sometimes, the
parton model results obtained with a particular set are compared with
$k_T$-factorization results based on another set.

In our previous paper \cite{our} we have presented a comparison of
results obtained with the help of $k_T$-factorization and the parton model.
The main goal of the paper \cite{our} was to demonstrate the differences
in the qualitative and numerical predictions coming from the matrix
elements. To simplify the calculations and to avoid various additional
dependences we had used gluon distribution which had only a reasonable
qualitative behaviour and a fixed value of $\alpha_S$.

The aim of this paper is to present a comparison between the
results of the conventional parton model and the $k_T$-factorization
approach for the quantities which are measured experimentally. For this
reason we use the realistic gluon distribution GRV94 \cite{GRV}
compatible with the most recent data, see discussion in Ref. \cite{GRV1}.

Below we shortly repeat the main formalism of the approaches used,
discuss the values of the parameters and present the numerical results
on charm and beauty production obtained in the LO (and qualitatively in NLO)
parton model and in the $k_T$-factorization approach.

\section{Conventional parton model approach}

The conventional parton model expression for the calculation of heavy
quark hadropro\-duc\-tion cross sections has the factorized form
\cite{CSS}:
\begin{equation}
\sigma (a b \rightarrow Q\overline{Q}) =
\sum_{ij} \int dx_i dx_j G_{a/i}(x_i,\mu_F) G_{b/j}(x_j,\mu_F)
\hat{\sigma} (i j \rightarrow Q \overline{Q}) \;,
\label{pm}
\end{equation}
where $G_{a/i}(x_i,\mu_F)$ and $G_{b/j}(x_j,\mu_F)$ are the structure
functions of partons $i$ and $j$ in the colliding hadrons $a$ and $b$,
$\mu_F$ is the factorization scale (i.e. virtualities of incident
partons) and $\hat{\sigma} (i j \rightarrow Q \overline{Q})$ is
the cross section of the subprocess which is calculated in
perturbative QCD. The latter cross section can be written as a sum of
LO and NLO contributions,
\begin{eqnarray}
\hat{\sigma} (i j \rightarrow Q\overline{Q}) &=&
\frac{\alpha_s^2(\mu_R)}{m_Q^2}\biggl(f^{(o)}_{ij}(\rho)
+ 4 \pi \alpha_s(\mu_R)
\Bigl[f_{ij}^{(1)}(\rho) + \bar f_{ij}^{(1)}(\rho)
\ln(\mu^2/m_Q^2)\Bigr]\biggr) \; ,
\label{shat}
\end{eqnarray}
where $\mu_R$ is the renormalization scale and $f^{(o)}_{ij}$ as
well as $f^{(1)}_{ij}$ and $\bar{f}^{(1)}_{ij}$ depend only on the single
variable
\begin{equation}
\rho = \frac{4m_Q^2}{\hat{s}} \;, \; \hat{s} = x_i x_j s_{ab} \;.
\end{equation}
(In the factor $\ln(\mu^2/m_Q^2)$ we assume $\mu_R = \mu_F$ following
\cite{1}. In the case of different values of $\mu_R$ and $\mu_F$, which
is preferable for the description of the experimental data \cite{FMNR},
Eq.~(2) becomes more complicated.)

The expression (1) corresponds to the process shown schematically in
Fig.~1a. The main contribution to the cross section at small $x$ is
known to come from gluons, $i = j = g$.

Usually in the parton model the values
\begin{equation}
\mu_F = \mu_R = m_Q
\end{equation}
are used for the total cross sections and
\begin{equation}
\mu_F = \mu_R = m_T = \sqrt{m_Q^2 + p_Q^2}
\end{equation}
for the one-particle distributions \cite{FMNR}. However we calculate
the total cross sections of heavy quark production as the integrals
over their $p_T$ distrubutions, i.e. with scales~(5).

Both in the parton model and in the $k_T$-factorization approach we take
\begin{equation}
m_c = 1.4\; {\rm GeV}, \qquad m_b = 4.6\; {\rm GeV} \;,
\end{equation}
for the values of short-distance perturbative quark masses
\cite{Nar,BBB}.

Another principal problem of the parton model is the collinear
approximation. The transverse momenta of the incident partons, $q_{iT}$
and $q_{jT}$ are assumed to be zero, and their virtualities are
accounted for through the structure functions only; the cross section
$\hat{\sigma} (i j \rightarrow Q \overline{Q})$ is assumed to be
independent of these virtualities. Naturally, this approximation
significantly simplifies the calculations.

The conventional NLO parton model approach with collinear approximati\-on
works quite reasonably for one-particle diustributions and for the total
cross sections; at the same time it is in serious disagreement with
the data on heavy quark correlations (without $k_T$ kick
introduction \cite{FMNR}).

\section{Heavy quark production in the $k_T$-factorization approach}

In the $k_T$-factorization approach the transverse momenta of the
incident gluons in the small-$x$ region result from the
$\alpha_s \ln k_T^2$ diffusion in the gluon evolution. The diffusion is
described by the function $\varphi(x,q^2)$ giving the gluon distribution
at a fixed fraction of the longitudinal momentum of the initial hadron,
$x$, and of the gluon virtuality, $q^2$. At very low $x$ that is to
leading $\log(1/x)$ accuracy, it is approximately determined \cite{GLR}
via the derivative of the usual structure function:
\begin{equation}
\label{xG}
\varphi (x,q^2)\ =\ 4\sqrt2\,\pi^3 \frac{\partial[xG(x,q^2)]}
{\partial q^2}\ .
\end{equation}
Such a definition of $\varphi(x,q^2)$ enables to treat correctly
the effects arising from the gluon virtualities.

Although generally $\varphi$ is a function of three variables, $x$,
$q_T$ and $q^2$, the transverse momentum dependence is comparatively
weak since $q_T^2 \approx - q^2$ for small $x$ in LLA in agreement with
$q^2$-dependences of structure function. Note that due to QCD scaling
violation the value $\varphi(x,q^2)$ for the realistic structure
functions increases more fast with decreasing of $x$. Therefore at
smaller $x$, larger $q_T$ becomes important in the numerical calculations.

The exact expression of the $q_T$ gluon distribution can be obtained as a
solution of the evolution equation which, contrary to the parton model
case, is nonlinear due to interactions between the partons in the small
$x$ region. The calculations \cite{Blu} of the $q_T$ gluon distribution
in leading order using the BFKL theory \cite{BFKL} result in differences
from our $\varphi(x,q^2)$ function given by Eq. (7) by only about 10-15\%.

Here we deal with the matrix element accounting for the gluon
virtualities and polarizations. Since it is much more complicated than
in the parton model we consider only the LO of the subprocess
$gg \to Q\bar{Q}$ which gives the main contribution to the heavy quark
production cross section at small $x$, see the diagrams in Fig.~2. The
lower and upper ladder blocks present the two-dimensional gluon
functions $\varphi(x_1,q_1^2)$ and $\varphi(x_2,q_2^2)$.

Strictly speaking, Eq. (7) may be justified in the leading $\log (1/x)$
limit only. To restore the unintegrated parton distribution $f_a(x,q_T,\mu)$
(i.e. the probability to find a parton $a$ with transverse momentum $q_t$
which initiates our hard process, with factorization scale $\mu$)
based on the conventional (integrated) parton density
$a(x,\lambda^2)$ we have to consider the DGLAP
evolution\footnote{For the $g \to gg$ splitting we need to insert a
factor $z'$ in the last integral of Eq. (8) to account for the identity
of the produced gluons.}
\begin{equation}
\frac{\partial a}{\partial \ln \lambda^2} = \frac{\alpha_s}{2 \pi}
\left[ \int^{1-\Delta}_x P_{aa'} (z) a'(\frac xz,\lambda^2) dz -
a(x,\lambda^2) \sum_{a'} \int^{1-\Delta}_0 P_{a'a} (z') dz' \right]
\end{equation}
(here $a(x,\lambda^2)$ denotes $xg(x,\lambda^2)$ or $xq(x,\lambda^2)$
and $P_{aa'}(z)$ are the splitting functions).

The first term on the right-hand side of Eq. (8) describes the number
of partons $\delta_a$ emitted in the interval
$\lambda^2 < q^2_t < \lambda^2+\delta \lambda^2$, while the second
(virtual) term reflects the fact that the parton $a$ disappears after the
splitting.

The second contribution may be resummed to give the survival probability
$T_a$ that the parton $a$ with transverse momentum $q_t$ remains
untouched in the evolution up to the factorization scale
\begin{equation}
\label{Sud}
T_a(q_t,\mu) = \exp \left[ -\int^{\mu^2}_{q^2_t}
\frac{\alpha_s(p_t)}{2\pi} \frac{dp^2_t}{p^2_t}
\sum_{a'} \int^{1-\Delta}_0 P_{a'a} (z') dz' \right]
\end{equation}
Thus the unintegrated distribution $f_a(x,q_t,\mu)$ reads
\begin{equation}
f_a(x,q_t,\mu) = \left[\frac{\alpha_s}{2 \pi} \Theta (1-\delta-x)
\int^{1-\Delta}_x P_{aa'} (z) a'\left(\frac xz,q_t^2\right) dz \right]
T_a(q_t,\mu) \;,
\end{equation}
where the cut-off $\delta = q_t/\mu$ is used both in Eqs. (9) and (10)
\cite{MW1,KMR}.

In the leading $\log (1/x)$ (i.e. BFKL) limit the virtual DGLAP
contribution is neglected. So $T_a = 1$ and one comes back to Eq. (7)
\begin{equation}
f_a^{BFKL} (x,q_t,\mu) = \frac{\partial a(x,\lambda^2)}
{\partial \ln \lambda^2}\;\;\;, \lambda^{\epsilon} = q_t \; .
\end{equation}

In the double log limit Eq. (10) can be written in the form
\begin{equation}
f_a^{DDT}(x,q_t,\mu) = \frac\partial{\partial \ln \lambda^2}
\left[a(x,\lambda^2) T_a(\lambda,\mu)\right]_{\lambda = q_t} \; ,
\end{equation}
which was firstly proposed by \cite{DDT}. In this limit the
derivative $\partial T_a /\partial \ln \lambda^2$ cancels the second
term of the r.h.s. of Eq. (8) (see \cite{KMR} for a more detailed
discussion).

Finally, the probability $f_a(x,q_t,\mu)$ is related to the BFKL
function $\varphi(x,q^2_t)$ by
\begin{equation}
\varphi (x,q^2)\ =\ 4\sqrt2\,\pi^3 f_a(x,q_t,\mu) \; .
\end{equation}

Note that due to a virtual DGLAP contribution the derivative
$\partial a(x,\lambda^2) / \partial \lambda^2$ can be negative for
not small enough $x$ values. This shortcoming of Eq. (11) is overcome
partly in the case of Eq. (12). Unfortunately the cut-off $\Delta$
used in a conventional DGLAP computation does not depend on the scale
$\mu$. To obtain an integrated parton distributions it is enough to put
any small $\Delta \ll 1$\footnote{There is a cancellation between the
real and virtual soft gluon DL contributions in the DGLAP equation, written
for the integrated partons (including all $k_t\le\mu$). The emission of a
soft gluon with momentum fraction $(1-z)\to 0$ does not affect the
$x$-distribution of parent partons. Thus the virtual and real
contributions originated from $1/(1-z)$ singularity of the splitting
function $P(z)$ cancel each other.

On the contrary, in the unintegrated case the emission of soft gluon
(with $q'_t>k_t)$ alters the transverse momentum of parent ($t$-channel)
parton. Eq.~(8) includes this effect through the derivative
$\partial T(k^2_t,\mu^2)/\partial k^2_t$.}.

On the other hand, in the survival probability Eq. (9) we have to use the
true (within the leading log approximation) value $\Delta = q_t/\mu$.
Thus for a rather large $q_t$ (of the order of $\mu$) and $x$ even the DDT
form Eq. (12) is not precise enough. Only the expression (10) with the
same cut-off $\Delta$ in a real DGLAP contribution and in survival
probability (9) provides the positivity of the unintegrated probability
$f_a(x,q_t,\mu)$ in the whole interval $0 < x < 1$.

Of course, just by definition $f_a(x,q_t,\mu) = 0$ when the
transverse momentum $q_t$ becomes larger than the factorization scale
$\mu$.








In what follows we use the Sudakov decomposition for quark momenta
$p_{1,2}$ through the momenta of colliding hadrons  $p_A$ and
$p_B\,\, (p^2_A = p^2_B \simeq 0)$ and the transverse momenta
$p_{1,2T}$:
\begin{equation}
\label{1}
p_{1,2} = x_{1,2} p_B + y_{1,2} p_A + p_{1,2T}.
\end{equation}
The differential cross section of heavy quark hadroproduction has the
following form:\footnote{We put the argument of $\alpha_S$ to be equal to gluon
virtuality, which is very close to the BLM scheme\cite{blm}; (see also
\cite{lrs}).}
\begin{eqnarray}
\frac{d\sigma_{pp}}{dy^*_1 dy^*_2 d^2 p_{1T}d^2
p_{2T}}\,&=&\,\frac{1}{(2\pi)^8}
\frac{1}{(s)^2}\int\,d^2 q_{1T} d^2 q_{2T} \delta (q_{1T} +
q_{2T} - p_{1T} - p_{2T}) \nonumber \\
\label{spp}
&\times &\,\frac{\alpha_s(q^2_1)}{q_1^2} \frac{\alpha_s (q^2_2)}{q^2_2}
\varphi(q^2_1,y)\varphi (q^2_2, x)\vert M_{QQ}\vert^2.
\end{eqnarray}
Here $s = 2p_A p_B\,\,$, $q_{1,2T}$ are the gluon transverse momenta
and $y^*_{1,2}$  are the quark rapidities in the hadron-hadron c.m.s.
frame,
\begin{equation}
\label{xy}
\begin{array}{lcl}
x_1=\,\frac{m_{1T}}{\sqrt{s}}\, e^{-y^*_1}, &
x_2=\,\frac{m_{2T}}{\sqrt{s}}\, e^{-y^*_2},  &  x=x_1 + x_2\\
y_1=\, \frac{m_{1T}}{\sqrt{s}}\, e^{y^*_1}, &  y_2 =
\frac{m_{2T}}{\sqrt{s}}\, e^{y^*_2},  &  y=y_1 + y_2 \\
&m_{1,2T}^2 = m_Q^2 + p_{1,2T}^2. &
\end{array}
\end{equation}
$\vert M_{QQ}\vert^2$ is the square of the matrix element for the heavy
quark pair hadropro\-duction.

In LLA kinematic
\begin{equation}
\label{q1q2}
\begin{array}{crl}
q_1 \simeq \,yp_A + q_{1T}, & q_2 \simeq \,xp_B + q_{2T}.
\end{array}
\end{equation}
so
\begin{equation}
\label{qt}
\begin{array}{crl}
q_1^2 \simeq \,- q_{1T}^2, & q_2^2 \simeq \,- q_{2T}^2.
\end{array}
\end{equation}
(The more accurate relations are $q_1^2 =- \frac{q_{1T}^2}{1-y}$,
$q_2^2 =- \frac{q_{2T}^2}{1-x}$ but we are working in the kinematics
where $x,y \sim 0$).

The matrix element $M$ is calculated in the Born approximation of QCD
(see details in \cite{3,our} without standard simplifications of the
parton model.

\section{Total cross sections and one-particle distributions}

Eq.~(15) enables us to calculate straightforwardly all distributions
concerning one-particle or pair production. One-particle
calculations as well as correlations between two produced heavy quarks
can be easely obtained using, say, the VEGAS code \cite{Lep}.

However there exists a principal problem coming from the infrared region.
Since the functions $\varphi (x,q^2_2)$ and $\varphi (y,q^2_1)$ are unknown
at small values of $q^2_2$ and $q^2_1$ we use the direct
consequence of Eq.~(7) \cite{Kwi}
\begin{equation}
\label{xgg}
xG(x,q^2) \,=\, xG(x,Q_0^2) + \frac{1}{4\sqrt{2}\,\pi^3}
\int^{q^2}_{Q_0^2} \varphi (x,q^2_1)\,dq_1^2.
\end{equation}
and rewrite the integrals in the Eq.~(15) as
\newpage
$$\int d^2 q_{1T} d^2 q_{2T} \delta (q_{1T} +
q_{2T} - p_{1T} - p_{2T}) \frac{\alpha_s(q^2_1)}{q_1^2}
\frac{\alpha_s (q^2_2)}{q^2_2} \varphi(q^2_1,y)\varphi (q^2_2, x)
\vert M_{QQ}\vert^2 = $$
\begin{equation}
\label{int}
 = (4\sqrt{2}\,\pi^3 \alpha_s (m^2_T) )^2 \, xG(x,Q^2_0)\, yG(y,Q^2_0)\,
T^2(Q_0^2,\mu^2)\,
\left (\frac{\vert M_{QQ} \vert^2}{q^2_1 q^2_2}
\right)_{q_{1,2}\rightarrow 0} \; +
\end{equation}
$$ + \; 4\sqrt{2}\,\pi^3 \alpha_s (m^2_T) xG(x,Q^2_0)\,T(Q_0^2,\mu^2)\,
\int^{\infty}_{Q^2_0} \, d q^2_{1T}\, \delta (q_{1T} - p_{1T} -
p_{2T})\,\, \times $$
$$ \times \, \frac{\alpha_s (q^2_1)}{q^2_1} \varphi(q^2_1,y)
\left (\frac{\vert M_{QQ} \vert^2}{q^2_2}
\right)_{q_2\rightarrow 0} \; + $$
$$ + \; 4\sqrt{2}\,\pi^3 \alpha_s (m^2_T)\, yG(y,Q^2_0)\,T(Q_0^2,\mu^2)\,
\int^{\infty}_{Q^2_0} \, d q^2_{2T}\, \delta
(q_{2T} - p_{1T} - p_{2T})\, \, \times $$
$$ \times \, \frac{\alpha_s (q^2_2)}{q^2_2} \varphi(q^2_2,x)
\left (\frac{\vert M_{QQ}
\vert^2}{q^2_1} \right)_{q_1 \rightarrow 0} \; + $$
$$ + \; \int^{\infty}_{Q^2_0} \, d^2 q_{1T}
\int^{\infty}_{Q^2_0} \, d^2 q_{2T}\,
\delta (q_{1T} + q_{2T} - p_{1T} - p_{2T}) \, \times $$
$$ \times \,\frac{\alpha_s(q^2_1)}{q_1^2} \frac{\alpha_s (q^2_2)}{q^2_2}
\varphi(q^2_1,y)\varphi (q^2_2, x) \vert M_{QQ}\vert^2 \;, $$
where the unintegrated gluon distributions are taken from Eqs.~(9) and
(10).

The first contribution in Eq.~(20), with averaging of the matrix element
over the directions of the two-dimensional vectors $q_{1T}$ and $q_{2T}$,
is exactly the same as the conventional LO parton model expression, with
QCD scales $\mu_R^2 = m^2_T$ and $ \mu_F^2 = Q_0^2$ multiplied by the
'survival' probability $T^2(Q_0^2,\mu^2)$ not to have transverse momenta
$q_{1,2,T} > Q_0$. The sum of the produced heavy quark momenta is exactly
zero here.

The next three terms contain the corrections to the parton model related
to the gluon polarizations, virtualities and transverse momenta in the
matrix element. The relative contribution of the corrections strongly
depends on the initial energy. If it is not high enough, the first term
in Eq.~(20) dominates, and all results coincide practically (after
accounting for Eq.~(19)) with the conventional LO parton model predictions.
In the case of very high energy the opposite situation takes place, the
first term in Eq.~(20) can be considered as a small correction and our
results differ from the conventional ones. It is necessary to note that
the absolute as well as the relative values of all terms in Eq.~(20)
strongly depend on the T-factor inclusion (i.e., when we use Eq.~(10)
or Eq.~(7)).

Before the numerical comparison it is necessary to note that the NLO
parton model actually results only in a normalization factor in the
case of one-particle distributions, the shapes of LO and LO+NLO
distributions are almost the same, see \cite{NDE,Beer,Beer1,MNR1}. This
means that we can calculate the K-factor
\begin{equation}
K = \frac{\sigma (LO) + \sigma (NLO)}{\sigma (LO)} \;,
\end{equation}
say, from the results for the total production cross sections, and
restrict ourselves only to LO calculations of $p_T$, or rapidity
distributions multiplying them by the K-factors.

The numerical values of the K-factors depend \cite{Liu} on the
structure functions used, quark masses, QCD scales and initial energy,
the dependence on the renormalization scale $\mu_R$ being especially
important. This seems to be evident, because the LO contribution is
proportional to $\alpha_s^2$, whereas the NLO contribution is
proportional to $\alpha_S^3$. However the more important dependence at
high energies, when small $\rho$ values dominate, comes from the
structure of Eq.~(2). At $\rho \to 0$ the functions $f^{(1)}_{gg}$ and
$\bar{f}^{(1)}_{gg}$ have constant limits \cite{1},
$f^{(1)}_{gg}(\rho \to 0) \approx 0.1$ and
$\bar{f}^{(1)}_{gg}(\rho \to 0) \approx -0.04$, so due to Eq.~(2)
the K factor at high energies depends strongly on the ratio
$\mu / m_Q$.

First of all let us present the role of the $T$-factors, Eq.~(9). In
Fig.~3 we show their values which were calculated as the ratios of the
values of the first term of Eq.~(20) to the same values calculated with
$T_a(q_{1t},\mu)=1$ for the cases of charm and beauty production at
$\sqrt s=14$~TeV and $\mu^2=\hat{s}$ as functions of $q_{1t}$. The values
of heavy quark transverse momenta were fixed at 20~GeV/c. In both cases
the values of $T_a(q_{1t})$ are rather small at small $q_{1t}$ and
$T_a(q_{1t})\to 1$ at $p_t \gg q_{1t}$.

Now let us compare the numerical results predicted by the parton model
and by the $k_T$-factorization approach.

The energy dependences of the total cross sections of $c\bar{c}$ and
$b\bar{b}$ pair production are presented in Fig.~4. As was mentioned, at
comparatively small energies the first term in Eq.~(20) dominates and
the results of the $k_T$-factorization approach should be close to the
LO parton model prediction. Actually the first results are even smaller
due to the presence of the $T$-factor in Eq.~(10). However the
$k_T$-factorization approach predicts a stronger energy dependence than
the LO parton model both for $c\bar{c}$ and $b\bar{b}$ production. This
can be explained by additional contributions appearing at very high
energies in the $k_T$-factorization approach, see \cite{our}

The calculated values of one-particle $p_T$ distributions,
$d \sigma /dp_T$, in the $k_T$-factorization approach and in the LO
parton model are presented in Fig.~5. In all cases the $k_T$-factorization
predicts broader distributions. The average values of $p_T$ of the
produced heavy quarks are rather different in these two approaches, as
one can see from Table 1.

\newpage
\vskip 10 pt
\begin{center}
{\bf Table 1}
The average values of charm and beauty quark transverse momenta
$<\!p_T\!>$ (in GeV/c) in the $k_T$-factorization approach with
$\mu2 = \hat{s}$ and in the LO parton model.
\vskip 20 pt
\begin{tabular}{|c|r|r|r|r|} \hline

 & \multicolumn{2}{c|}{LO parton model} &
\multicolumn{2}{c|}{$k_T$-factorization}  \\ \hline

$\sqrt{s}$ & $c\bar{c}$ & $b\bar{b}$ & $c\bar{c}$ & $b\bar{b}$
\\ \hline

14 TeV   & 1.78 & 4.53 & 2.23 & 5.47 \\ \hline

1.8 TeV  & 1.48 & 3.96 & 1.91 & 4.54  \\  \hline
\end{tabular}
\end{center}
\vskip 10 pt

It seems to be very natural, because, contrary to the case of the LO
parton model, a large $p_T$ of one heavy quark can be compensated not
only by the $p_T$ of another heavy quark but also by the initial gluons
(i.e. by hard gluon jet emission).

The rapidity distributions of produced heavy quarks presented in Fig.~6
show that the main part of the difference between the $k_T$-factorization
approach and the LO parton model comes from the central region.

\section{Two-particle correlations}

We saw from the previous section that there is only a small difference
in our results for the total cross sections and one-particle
distributions obtained in the $k_T$-factorization and in the LO parton
model. The predictions of the NLO parton model for these quantities
differ from the LO parton model only by a normalization factor of
2-2.5 \cite{NDE,Beer,Beer1,MNR1}. So the difference between our
predictions and the NLO parton model should be small.

The calculations of two-particle correlations in different approaches are
more informative. The simplest quantity here is the distribution in
the total transverse momentum of the produced heavy quark pair,
$p_{pair}$. In LO, evidently,
$p_{pair} = p_{1T} + p_{2T} = q_{1T} + q_{2T}$, and if
$q_{1T} = q_{2T} = 0$, then $d \sigma /d p_{pair}$ is a $\delta$-function
of zero. So the $p_{pair}$ distributions give direct information about
the transverse momentum distribution of the incident partons.

It is clear that if $q_{iT} \ll p_{iT}$, then the distributions in
$p_{pair}$ should be narrower in comparison with the one-particle
$p_T$ distributions. In this case the Weizsaecker-Williams approximation
should be valid and one can believe that the parton model reflects the
real dynamics of the interaction. In the opposite case,
$q_{iT} \sim p_{iT}$, the large transverse momentum of the produced
heavy quark can be compensated not by the other quark, but by a
high-$p_T$ gluon. We have shown in our previous paper \cite{our}, that
about 70-80\% of the total cross section of high-$p_T$ quark production
at high energies originates from such processes, when the heavy quark
propagator is close to the mass shell.

We calculate the values of $d \sigma /d p_{pair}$ for charm (a) and
beauty (b) production in the $k_T$-factorization approach using the
unintegrated gluon distribution Eqs.~(9), (10) with scale values
$\mu^2 = \hat{s}$ and $\mu^2 = \hat{s}/4$ (only for $\sqrt s =14$~Tev).
Our results for pair production at different initial energies are shown
by solid curves in Fig.~7. For the comparison we present by dashed
curves the one-particle $p_T$-distributions taken from Fig.~5, obtained
in the same $k_T$-factorization approach and with the same $T$-factor.
As we put $Q_0^2$ = 1 GeV$^2$ in Eq.~(20), we can not distinguish
between the initial gluons with $q_T$ equal to, say, 0.1 GeV/c and
0.9 GeV/c, so our first bin in the $d \sigma /d p_{pair}$ distribution
has the width 2~GeV/c which explains some irregular behaviour of the
solid curves at small $p_T$. Naturally, all the solid and dashed curves
are normalized equally at the same energy.

At comparatively small energies, $\sqrt{s}$ = 39~GeV and even at
$\sqrt{s}$ =630~GeV the distributions $d \sigma /d p_{pair}$ are
narrower than the one-particle distributions $d \sigma /d p_T$. This
means that the transverse momenta of the produced heavy quarks almost
completely compensate each other. However the situation changes
drastically with increasing of the initial energy. Starting from
comparatively small $p_T$, the difference between the curves decreases
with energy. At $\sqrt{s}$ = 14 TeV the distributions are similar
both in the cases of $c\bar{c}$ and $b\bar{b}$ production. This means
that the production mechanism changes in the discussed energy region.
At $\sqrt{s}$ = 14~TeV the transverse momentum of the produced heavy
quark is balanced more probably by one or several gluons (see also
\cite{our}), because the contribution with large virtuality in the quark
propagator is more suppressed in comparison with the large virtuality in
the gluon propagator.

The discussed behaviour depends on the value of the scale $\mu^2$ in the
$T$-factor, Eq. (9). The similar calculation at energy $\sqrt{s}$
= 14 TeV with $\mu^2 = \hat{s}/4$ is shown in Fig.~7 for pair and single
production by dash-dotted and dotted curves, respectively. Here the
difference between these two curves is more significant and it becomes
larger for lower energies.

The distributions of the produced heavy quark pair as a function of
the rapidity gap $\Delta y = \vert y_Q - y_{\bar{Q}}\vert$ between
quarks are presented in Figs.~8. Here the difference between the LO PM
and the $k_T$-factorization predictions is not large again except for
the region of very large $\Delta y$.

Another interesting correlation is the distribution in the azimuthal
angle $\phi$, which is defined as the opening angle between the two
produced heavy quarks, projected onto the plane perpendicular to the beam
and defined as the $xy$-plane. In the LO parton model the sum of the
produced heavy quark momenta projected onto this plane is exactly zero,
and the angle between them is always 180$^o$. In the case of the NLO
parton model the $\phi$ distribution is non-trivial \cite{MNR},
however the predicted distribution (without including the $k_T$ kick) is
narrower in comparison with the fixed target data \cite{FMNR}.

The theoretical as well as experimental investigation of such
distributions are very important to control our understanding of the
considered processes. The problem is that in the case of one-particle
inclusive distributions for heavy quark production in hadron collisions
the sum of LO and NLO contributions of the parton model practically
coincides \cite{NDE,Beer,Beer1,MNR1} with the LO contribution multiplied
by K-factors. Therefore agreement with experimental data can be achieved
for too small or too large NLO contribution by fitting one parameter,
which can work as a normalization factor (say, the QCD scale). The
deviation in azimuthal correlations from the trivial $\delta(\phi - \pi)$
distribution comes from NLO correction to PM. However the standard NLO
contributions are not large enought for the description of the data and
only the comparatively large intrinsic transverse momentum of incoming
partons ($k_T$-kick) allows to describe \cite{FMNR} the data.

Preliminary results for the azimuthal correlations in the
$k_T$-factorization approach were considered in \cite{SS}. The main
difference in the information coming from $d \sigma /d p_{pair}$ and
$d \sigma/d \phi$ distributions is due to the comparatively slow heavy
quark. It gives a negligibly small contribution to the
$d \sigma /d p_{pair}$ in the first case, whereas in the second one
each quark contributes to the distribution $d \sigma/d \phi$ practically
independently of its momentum, so all corrections coming from quark
confinement, hadronization and resonance decay can be important.

As was discussed above, the first contribution in Eq.~(20) is the same
as the conventional LO parton model in which the angle between the
produced heavy quarks is always 180$^o$. However the angle between two
heavy hadrons can be slightly different from this value due to a
hadronization processes. To take this into account we assume that in
this first contribution the probability to find a hadron pair with
azimuthal angle $180^o -\phi$ is determined by the expression
\begin{equation}
w_1(\phi)\,=\,\frac{p_h}{\sqrt{p_h^2+p_t^2}} \;,
\end{equation}
where $p_h=0.2$~Gev/c is a transverse momentum in the azimuthal plane
coming from the hadronization process. The other contributions of Eq.~(20)
result in a more or less broad $\phi$ distribution so we neglect their
small modification due to hadronization.

The $k_T$-factorization approach predictions for the azimuthal
correlation of heavy quarks produced in $pp$ collisions are presented
in Fig.~9 and one can see that they change drastically when the
initial energy increases from fixed target to the collider region.

\section{Conclusion}

We have compared the conventional LO Parton Model (PM) and the
$k_T$-factorization approach for heavy quark hadroproduction
at collider energies using a realistic gluon (parton) distribution.
Both the transverse momenta and rapidity distributions have been
considered, as well as two-particle correlations, such as the
distribution over the rapidity gap between two heavy quarks, their
azimuthal correlations and distributions of the total transverse
momentum of the produced heavy quark pair, $p_{pair}$.

It has been shown in \cite{our} that the contribution of the domain
with strong $q_T$ ordering ($q_{1,2T} \ll m_T=\sqrt{m_Q^2+p_T^2}$)
coincides in $k_T$-factorization approach with the LO PM prediction.
Besides this a very numerically large contribution appears at high
energies in $k_T$-factorization approach in the region
$q_{1,2T}\ge m_T$. It kinematically relates to the events where the
transverse momentum of heavy quark $Q$ is balanced not by the momentum
of antiquark $\overline Q$ but by the momentum of the nearest gluon.

This configuration is associated with the NLO (or even NNLO, if both
$q_{1,2T} \ge m_T$) corrections in terms of the PM with fixed number of
flavours, i.e. without the heavy quarks in the evolution. Indeed, as
was mentioned in \cite{1}, up to 80\% of the whole NLO cross section
originates from the events where the heavy quark transverse momentum is
balanced by the nearest gluon jet. Thus the large "NLO" contribution,
especially at large $p_T$, is explained by the fact that the virtuality
of the $t$-channel (or $u$-channel) quark becomes small in the region
$q_T \simeq p_T$ and the singularity of the quark propagator
$1/(\hat{p} - \hat{q}) - m_Q)$ in the "hard" QCD matrix element,
$M(q_{1T},q_{2T},p_{1T},p_{2T})$, reveals itself.

The double logarithmic Sudakov-type form factor $T$ in the
definition of unintegrated parton density (10) comprises
an important part of the virtual loop NLO (with respect to the PM)
corrections. Thus we demonstrate that $k_T$-factorization approach
collects already at the LO the major part of the contributions which play
the role of the NLO (and even NNLO) corrections to the conventional PM.
Therefore we hope that a higher order (in $\alpha_S$) correction to
the $k_T$-factorization would be rather small.

Another advantage of this approach is that a non-zero transverse
momentum of $Q \overline Q$-system
($p_{T pair}=p_{1T}+p_{2T} = q_{1T}+q_{2T}$) is naturally achieved in the
$k_T$-factorization. We have calculated the $p_{T pair}$ distribution
and compared it with the single quark $p_T$ spectrum. At
low energies the typical values of $p_{T pair}$ are much lesser than heavy
quark $p_T$ in accordance with collinear approximation. However
for LHC energy both spectra become close to each other indicating that
the transverse momentum of second heavy quark is relatively small. The
typical value of this momentum ($p_{T pair}=k_T$-kick) depends on the
parton structure functions/densities. It increases with the initial
energy ($k_T$-kick increases with the decreasing of the momentum
fractions $x,y$ carried by the incoming partons) and with the
transverse momenta of heavy quarks, $p_T$. Thus one gets a possibility
to describe a non-trivial azimuthal correlation without introducing a
large "phenomenological" intrinsic transverse momentum of the partons.

It is necessary to note that the essential values of the parton
transverse momenta $q_{1T}$ and $q_{2T}$ increase in our calculations
with the growth of the value of $p_T^{min}$ of detected $b$-quark. In
the language of $k_T$-kick it means that the values of
$\langle k_T^2\rangle$ also increase.

A more detail study of the heavy quark correlations in the
$k_T$-factorization approach, the role of DL factor $T(k_T^2,\mu^2)$
and the value of scale $\mu$ in $T$-factor will be published elsewhere.

\subsection*{Acknowledgements}

The presented calculations were carried out in ICTP.
One of us (Y.M.S) is grateful to Prof. S.Randjbar-Daemi for providing
this possibility and to the staff for creating good working conditions.
We are grateful to Yu.L.Dokshitzer, G.P.Korchemsky and M.N.Mangano
for discussions.\\
This work is supported by grants NATO OUTR.LG 971390 and
RFBR 98-02-17629.

\newpage

{\bf Figure captions}

\vspace{.5cm}

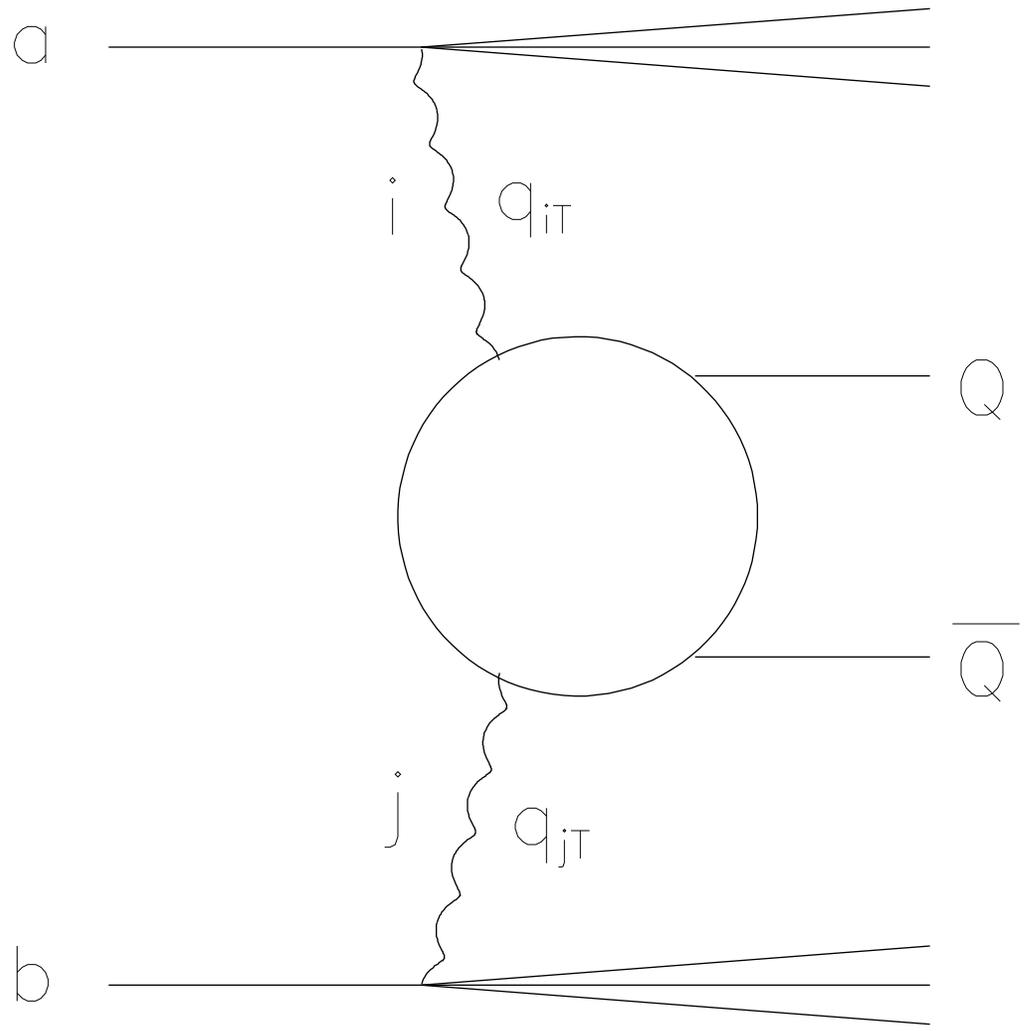
Fig. 1. Heavy quark production in hadron-hadron collision. The LO parton
model corresponds to the case when $q_{1T}=q_{2T}=0$.

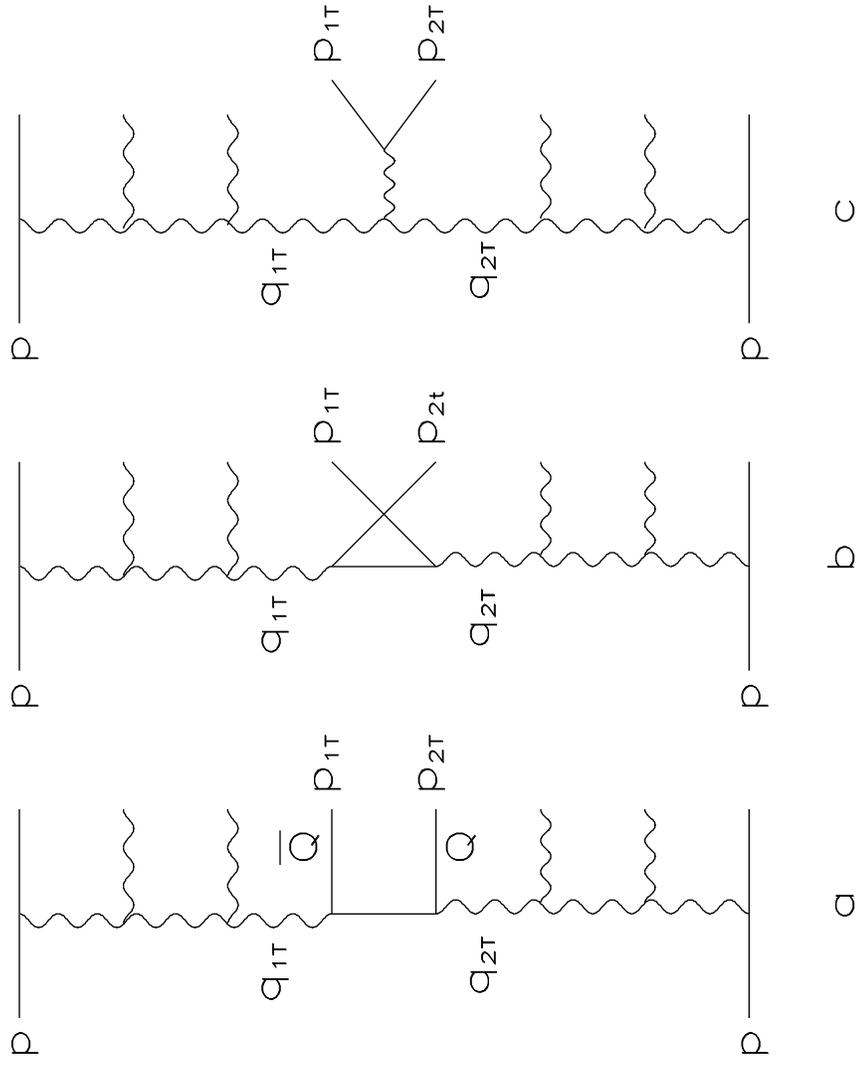
Fig. 2. Low order QCD diagrams for heavy quark production in $pp$
($p\overline{p}$) collisions via gluon-gluon fusion (a-c).

Fig. 3 The role of $T$-factor, Eq. (9) in the calculation of charm
(solid curve) and beauty (dashed curve) production with $p_T=20$~GeV/c
at $\sqrt{s}$ = 14 TeV.

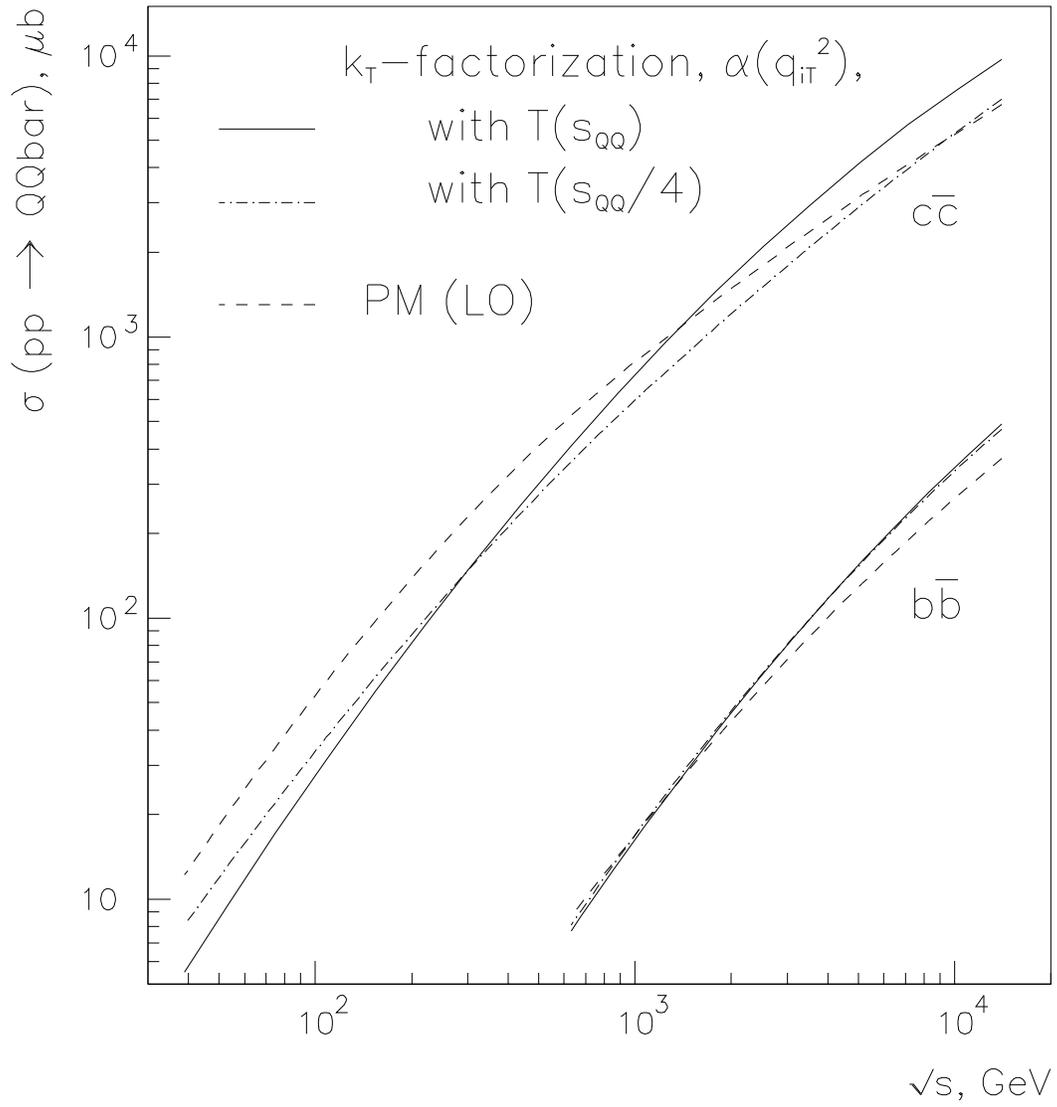
Fig. 4. The total cross sections for charm and beauty hadroproduction
in the $k_T$-factorization approach with unintegrated gluon distribution
$f_g(x,q_T,\mu)$ given by Eq.~(10), for $\mu^2$ values in Eq.~(9) equal
to $\hat{s}$ (solid curves), and $\hat{s}/4$ (dash-dotted curves), and
in the LO parton model (dashed curves).

Fig. 5. $p_T$-distributions of $c$-quarks (a) and $b$-quarks (b)
produced at different energies. Dashed curves are the results of the LO
parton model. Solid curves are calculated with unintegrated gluon
distribution $f_g(x,q_T,\mu)$ given by Eq.~(10), for
$\mu^2$ values in Eq.~(9) equal to $\hat{s}$ and dash-dotted curves
are calculated for $\mu^2 = \hat{s}/4$.

Fig. 6. Rapidity distributions of $c$-quarks (a) and $b$-quarks (b)
produced at different energies. Dashed curves are the results of the LO
parton model. Solid curves are calculated with the unintegrated gluon
distribution $f_g(x,q_T,\mu)$ given by Eqs.~(10), for
$\mu^2$ values in Eq.~(9) equal to $\hat{s}$ and dash-dotted curves
are calculated for $\mu^2 = \hat{s}/4$.

Fig. 7. The distributions of the total transverse momentum $p_{pair}$
for $c$-quarks (a) and $b$-quarks (b) produced
at different energies (solid curves), calculated with the unintegrated gluon
distribution $f_g(x,q_T,\mu)$ given by Eq.~(10) and~(10), for $\mu^2$
values in Eq.~(9) equal to $\hat{s}$. Dashed curves show the
one-particle (single) $p_T$-distributions with the same $\mu^2$ taken
from Fig.~5. Dash-dotted and dotted curves are the same calculations
for pair and single production at $\sqrt{s}$ = 14 TeV with
$\mu^2 = \hat{s}/4$.

Fig. 8. The calculated distributions of the rapidity difference between
two $c$-quarks (a) or $b$-quarks (b) produced at different energies in the
$k_T$ factorisation approach, calculated with unintegrated gluon
distribution $f_g(x,q_T,\mu)$ given by Eq.~(10), for $\mu^2$
values in Eq.~(9) equal to $\hat{s}$ (solid curves) and
$\mu^2 = \hat{s}/4$ (dash-dotted curves). Dashed curves show the LO PM
predictioms.

Fig. 9. The calculated azimuthal correlations of charm (a) and beauty
(b) pair production for $\mu^2=\hat{s}$ (solid curves) and
$\mu^2=\hat{s}/4$ (dash-dotted curves) at the energies equal to
$\sqrt{s}$ = 14 TeV, 1.8 TeV, 630 GeV and 39 GeV (the latter one only
for charm production).


\newpage

\begin{thebibliography}{99}

\bibitem{1} P.Nason, S.Dawson and R.K.Ellis. Nucl.Phys. B303 (1988) 607.
\bibitem{2} G.Altarelli et al. Nucl.Phys. B308 (1988) 724.
\bibitem{NDE} P.Nason, S.Dawson and R.K.Ellis. Nucl.Phys. B327 (1989)
49.
\bibitem{Beer} W.Beenakker,H.Kuijf, W.L.Van Neerven and J.Smith.
Phys.Rev. D40 (1989) 54.
\bibitem{Beer1} W.Beenakker, W.L.Van Neerven, R.Meng, G.A.Schuler
and J.Smith. Nucl.Phys. B351 (1991) 507.
\bibitem{MNR} M.N.Mangano, P.Nason and G.Ridolfi. Nucl. Phys. B373
(1992) 295.
\bibitem{FMNR} S.Frixione, M.N.Mangano, P.Nason and G.Ridolfi. Preprint
CERN-TH/97-16 (1997); hep-ph/9702287.
\bibitem{Shab} Yu.M.Shabelski. Talk, given at HERA Monte Carlo Workshop,
27-30 April 1998, DESY, Hamburg; hep-ph/9904492.
\bibitem{CCH} S.Catani, M.Ciafaloni and F.Hautmann. Phys.Lett. B242
(1990) 97; Nucl.Phys. B366 (1991) 135.
\bibitem{CE} J.C.Collins and R.K.Ellis. Nucl.Phys. B360 (1991) 3.
\bibitem{MW} G.Marchesini and B.R.Webber. Nucl.Phys. B386 (1992) 215.
\bibitem{CH} S.Catani and F.Hautmann. Phys.Lett. B315 (1993) 475;
Nucl.Phys. B427 (1994) 475.
\bibitem{CC} S.Camici and M.Ciafaloni. Nucl.Phys. B467 (1996) 25;
Phys.Lett. B396 (1997) 406.
\bibitem{GLR} L.V.Gribov, E.M.Levin and M.G.Ryskin. Phys.Rep. 100
(1983) 1.
\bibitem{LR} E.M.Levin and M.G.Ryskin. Phys.Rep. 189 (1990) 267.
\bibitem{8} E.M.Levin, M.G.Ryskin, Yu.M.Shabelski and A.G.Shuvaev.
Sov.J.Nucl.Phys. 53 (1991) 657.
\bibitem{lrs} E.M.Levin, M.G.Ryskin, Yu.M.Shabelski and A.G.Shuvaev.
Sov.J.Nucl.Phys. 54 (1991) 867.
\bibitem{3} M.G.Ryskin, Yu.M.Shabelski and A.G.Shuvaev. Z.Phys. C69
(1996) 269.
\bibitem{SS} Yu.M.Shabelski and A.G.Shuvaev. Eur.Phys.J. C6 (1999)
313.
\bibitem{our} M.G.Ryskin, Yu.M.Shabelski and A.G.Shuvaev.
hep-ph/9907507.
\bibitem{GRV} M.Gluck, E.Reya and A.Vogt. Z.Phys. C67 (1995) 433.
\bibitem{GRV1} M.Gluck, E.Reya and A.Vogt. Eur.Phys.J C5 (1998) 461.
\bibitem{CSS} J.C.Collins, D.E.Soper and G.Sterman. Nucl.Phys. B308
(1988) 833.
\bibitem{Nar} S.Narison. Phys.Lett. B341 (1994) 73; hep-ph/9503234.
\bibitem{BBB} P.Ball, M.Beneke and V.M.Braun. Phys.Rev. D52 (1995) 3929.
\bibitem{Blu} J.Bl\"{u}mlein. Preprint DESY 95-121 (1995).
\bibitem{BFKL} E.A.Kuraev, L.N.Lipatov and V.S.Fadin. Sov.Phys.JETP 45
(1977) 199.
\bibitem{MW1} G.Marchesini and B.R.Webber. Nucl.Phys. B310 (1988) 461.
\bibitem{KMR} M.A.Kimber, A.D.Martin and M.G.Ryskin. Hep-ph/9911379.
\bibitem{DDT} Yu.L.Dokshitzer, D.I.Dyakonov and S.I.Troyan. Phys.Rep. 58
(1980) 270.
\bibitem{blm} S.J.Brodsky, G.P.Lepage and P.B.Mackenzie. Phys.Rev. D28
(1983) 228.
\bibitem{Lep} G.P.Lepage. J.Comp.Phys. 27 (1978) 192.
\bibitem{Kwi} J.Kwiechinski. Z.Phys. C29 (1985) 561.
\bibitem{MNR1} M.N.Mangano, P.Nason and G.Ridolfi. Nucl. Phys. B405
(1993) 507.
\bibitem{Liu} Liu Wenjie, O.P.Strogova, L.Cifarelli and
Yu.M.Shabelski. Phys.Atom.Nucl. 57 (1994) 844.
\end{thebibliography}
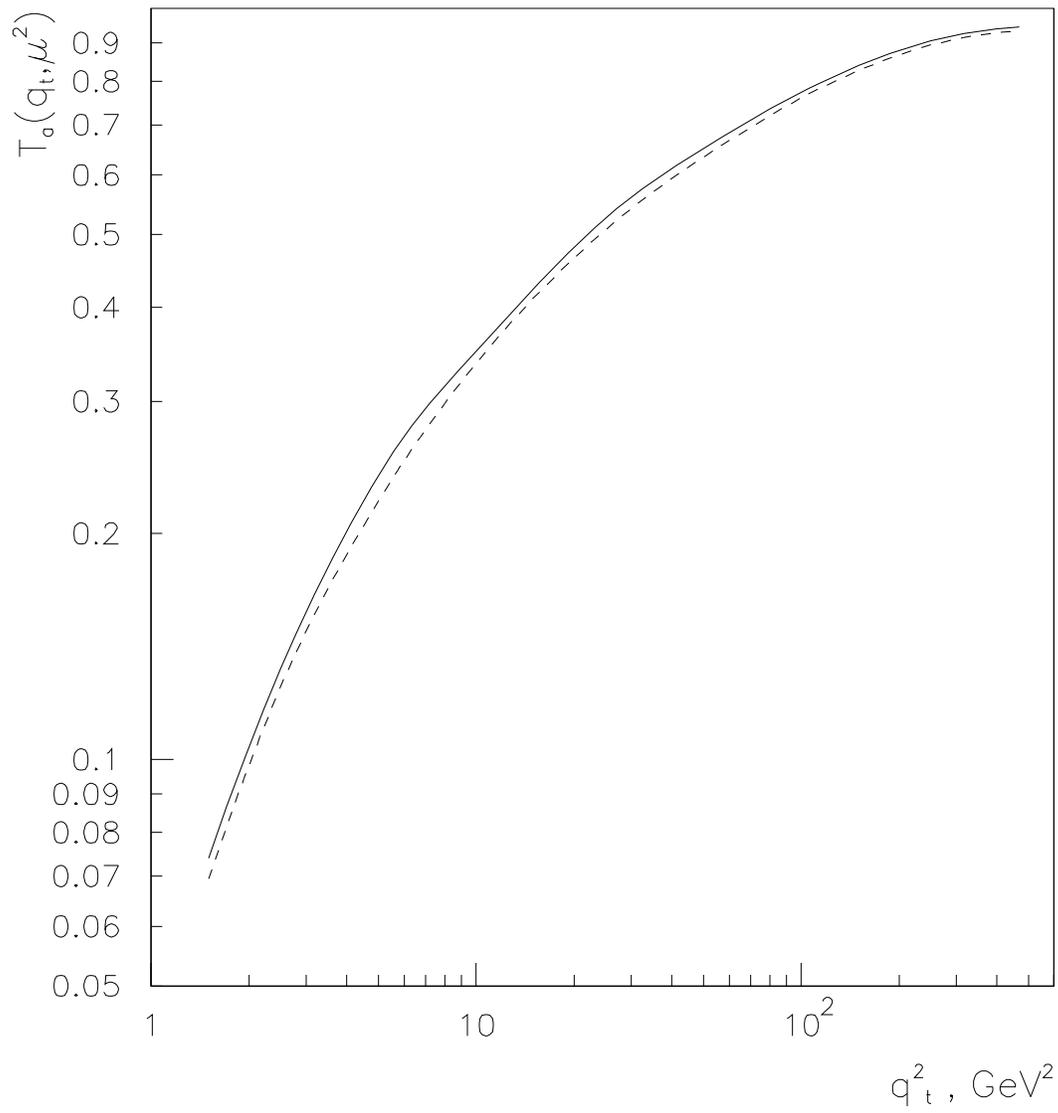
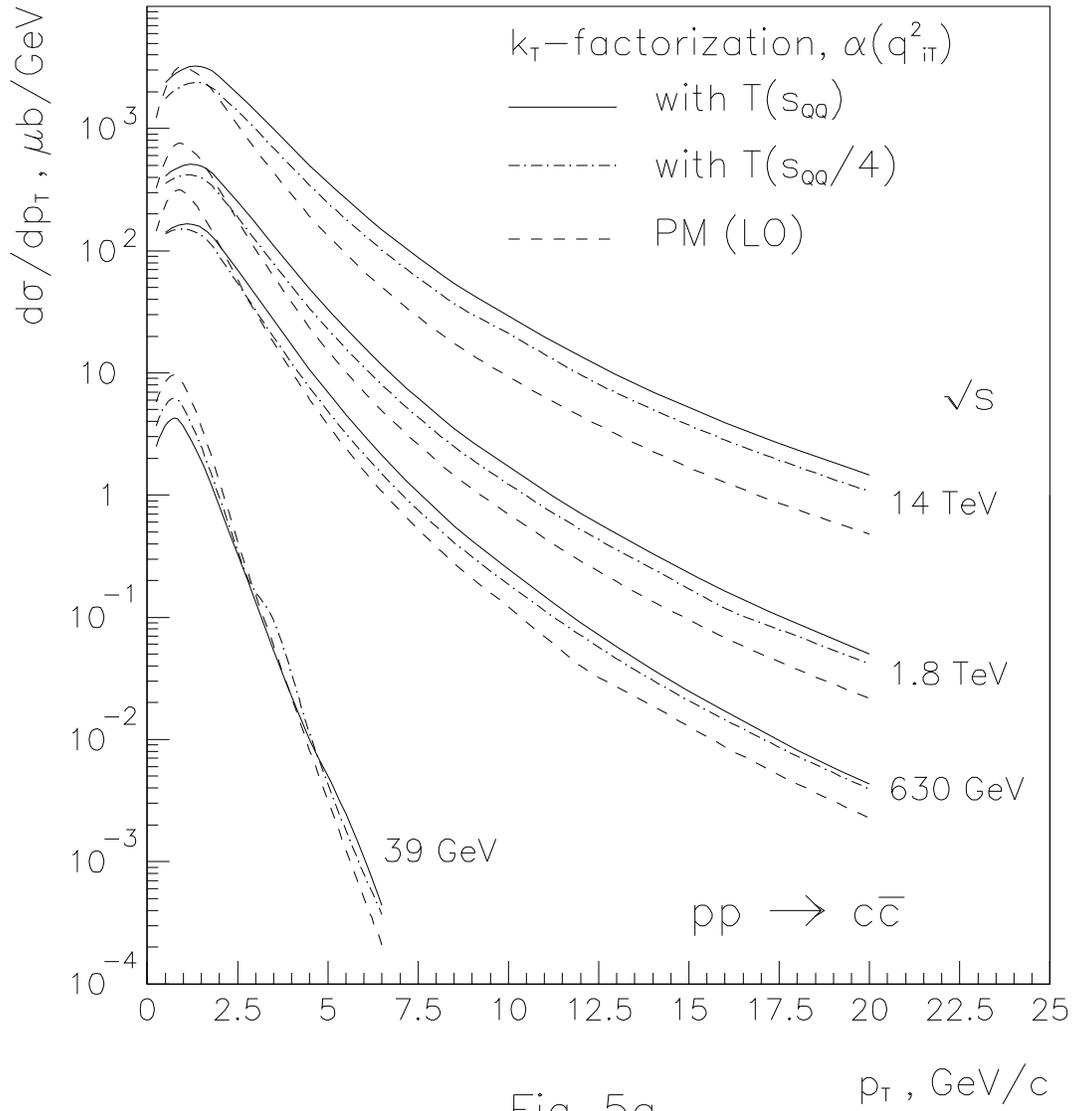
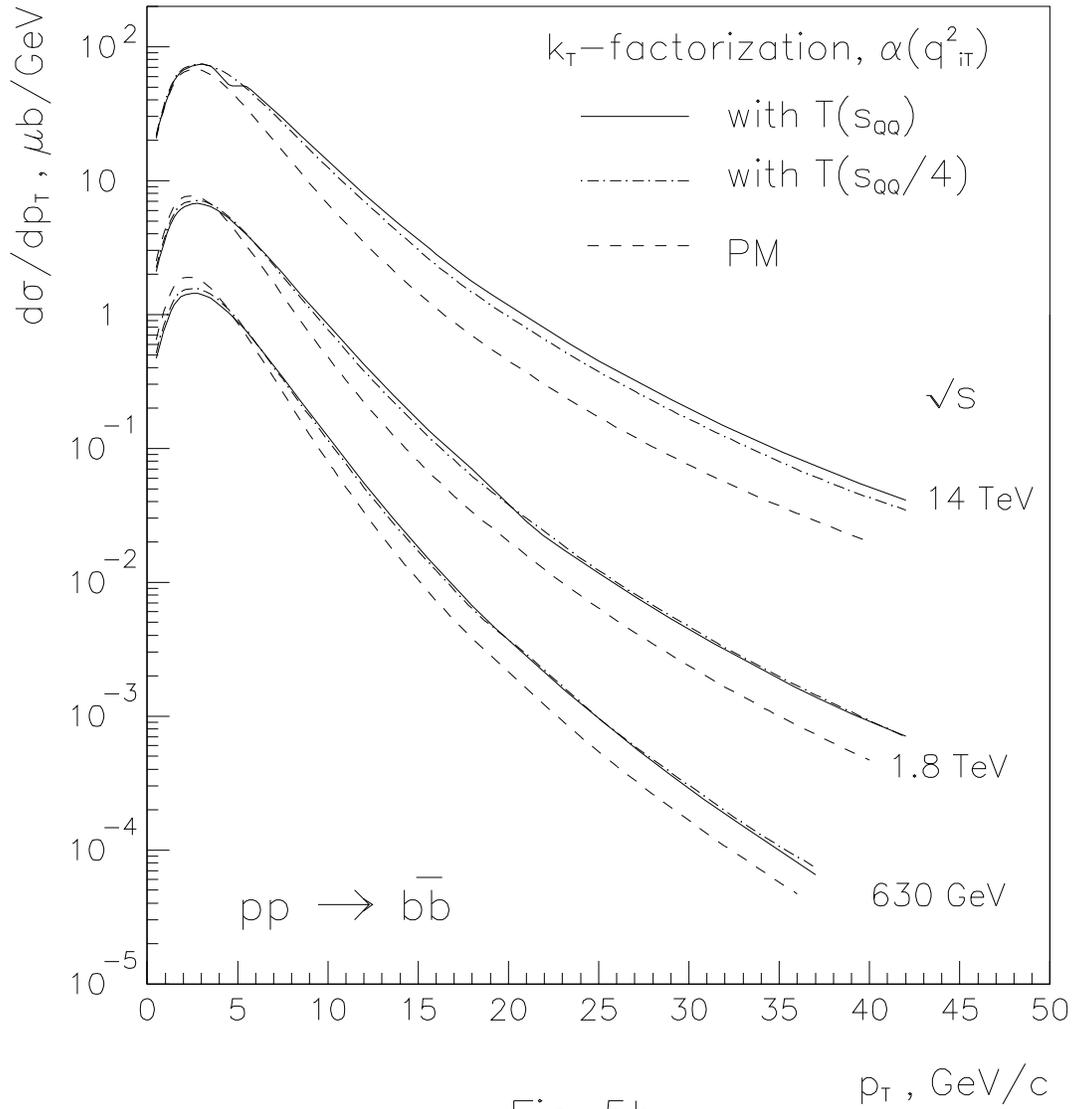
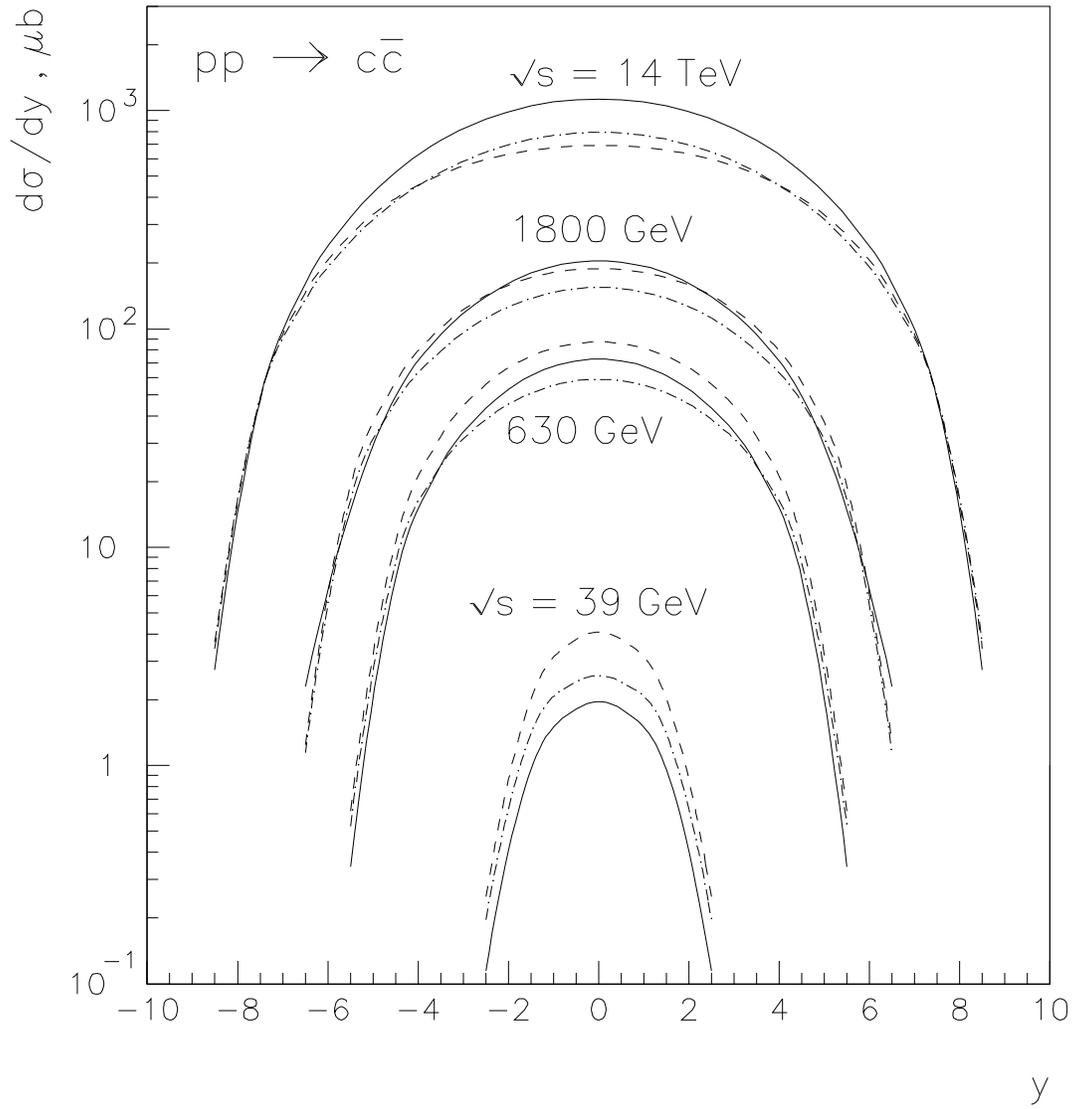
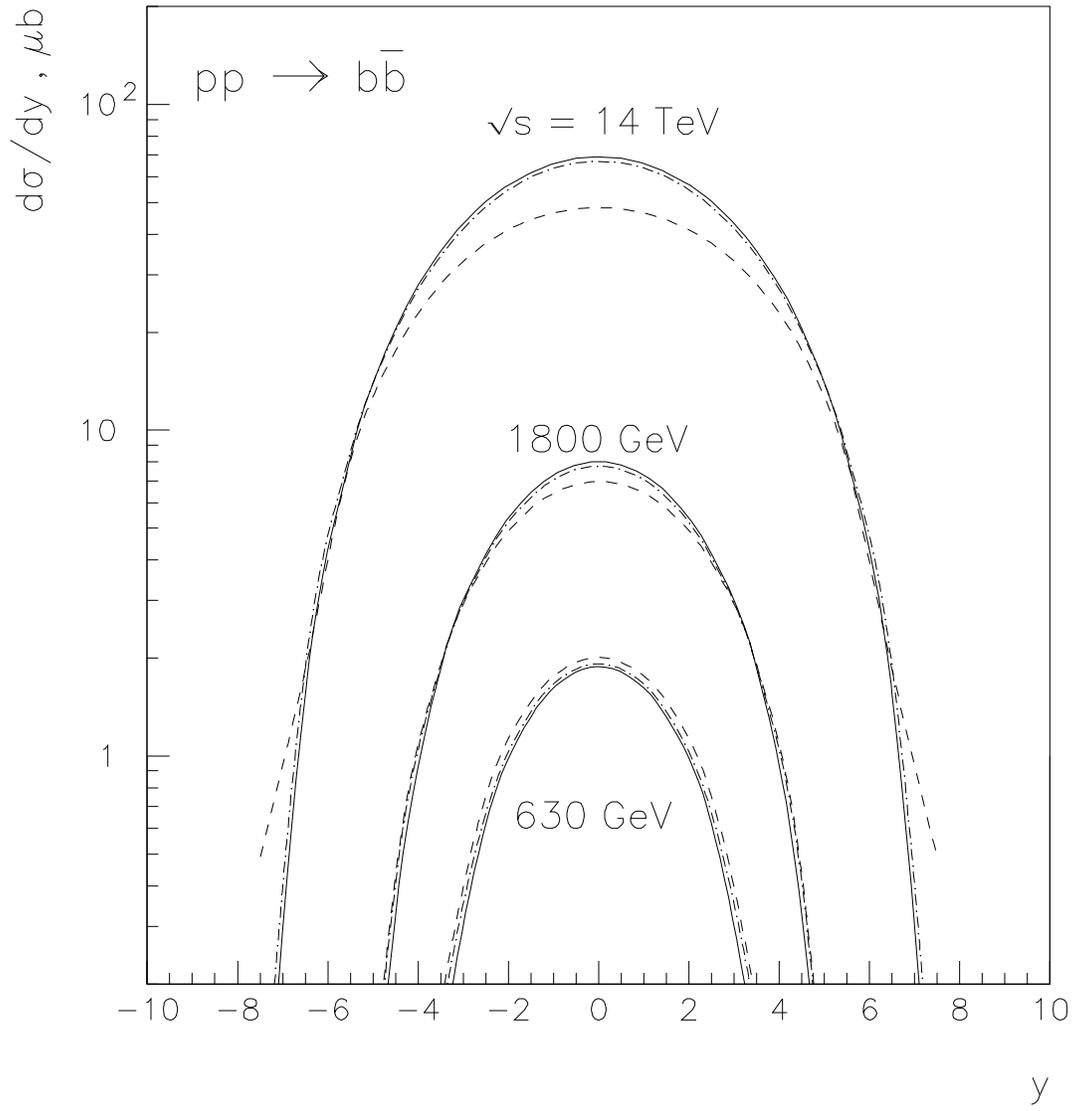
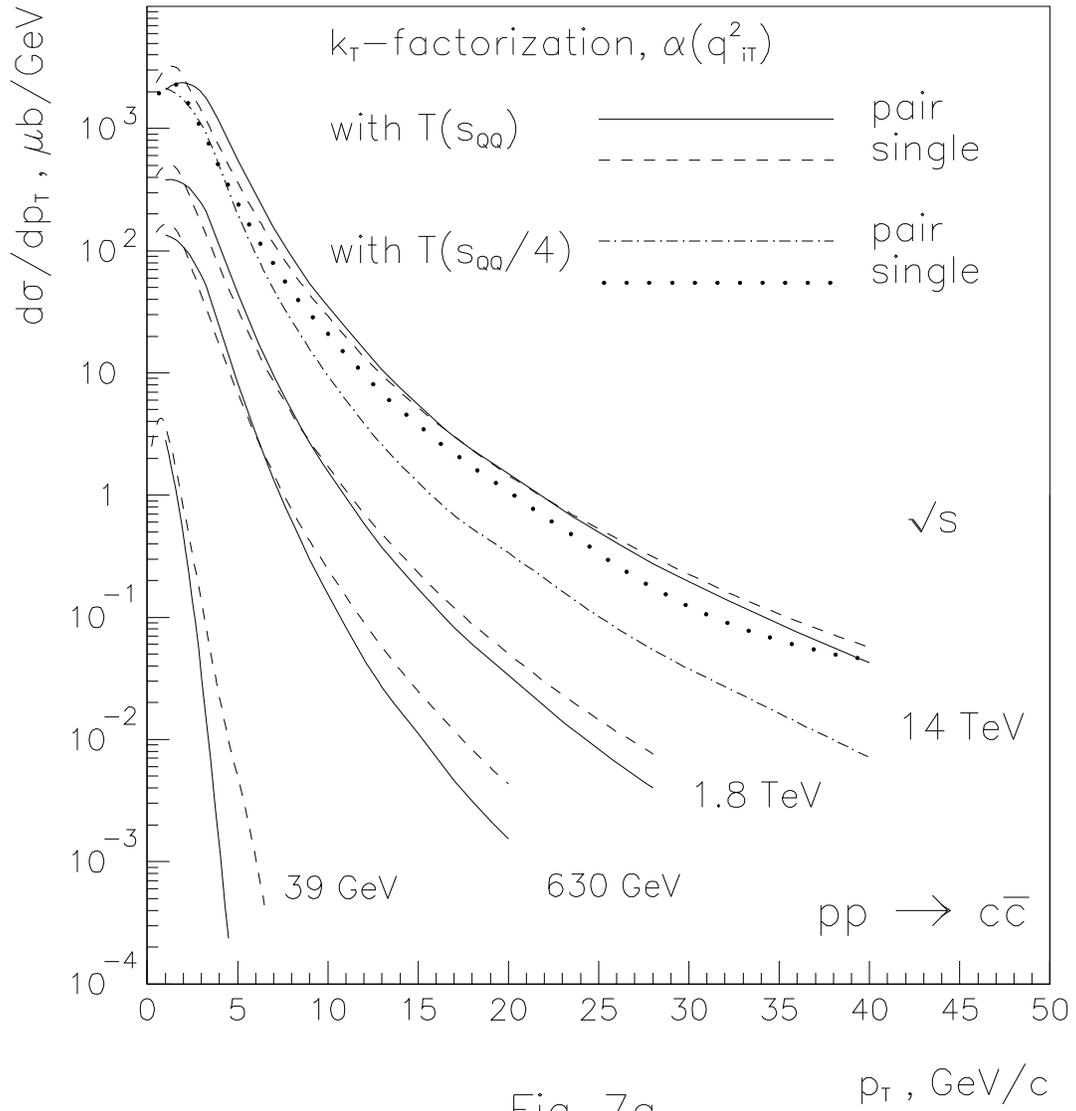
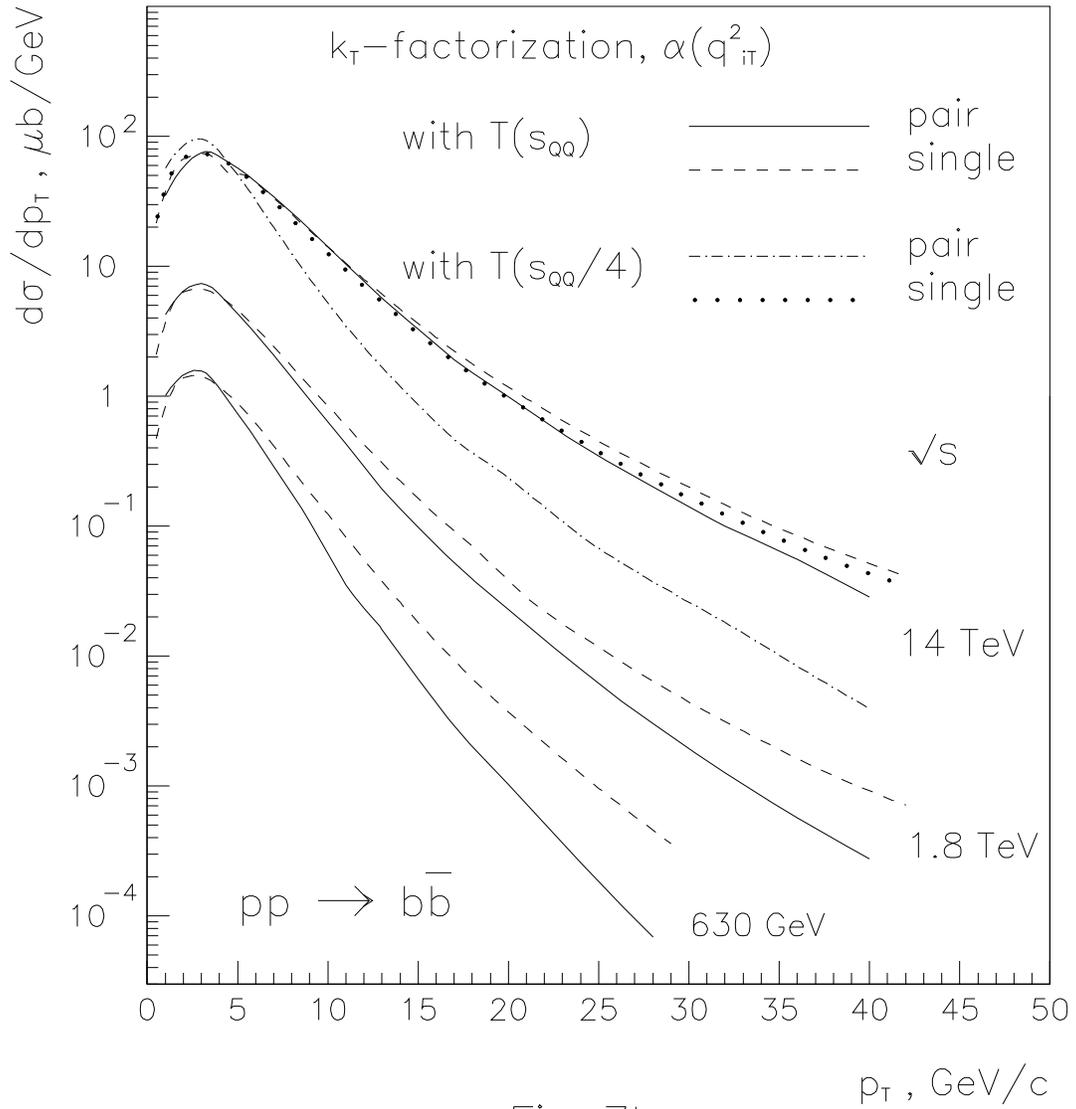
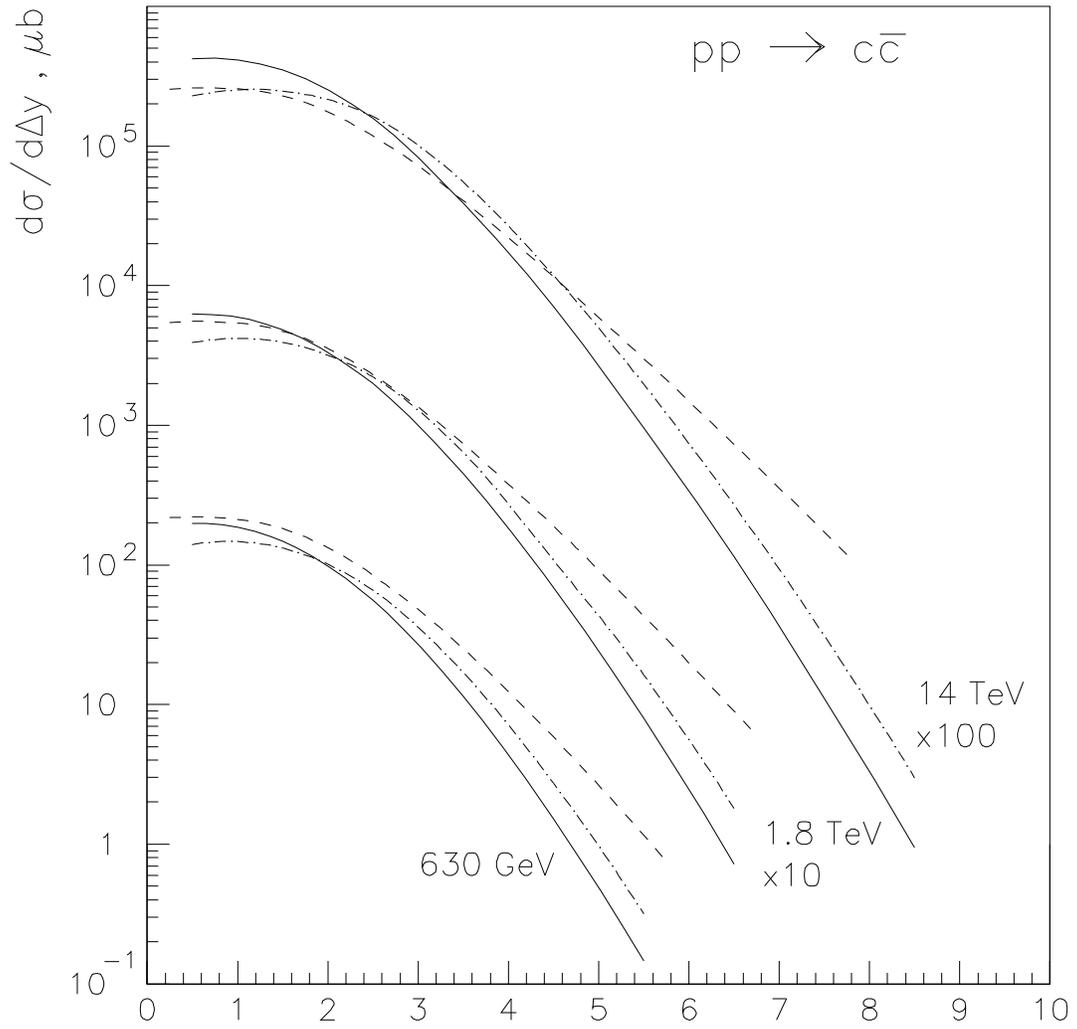
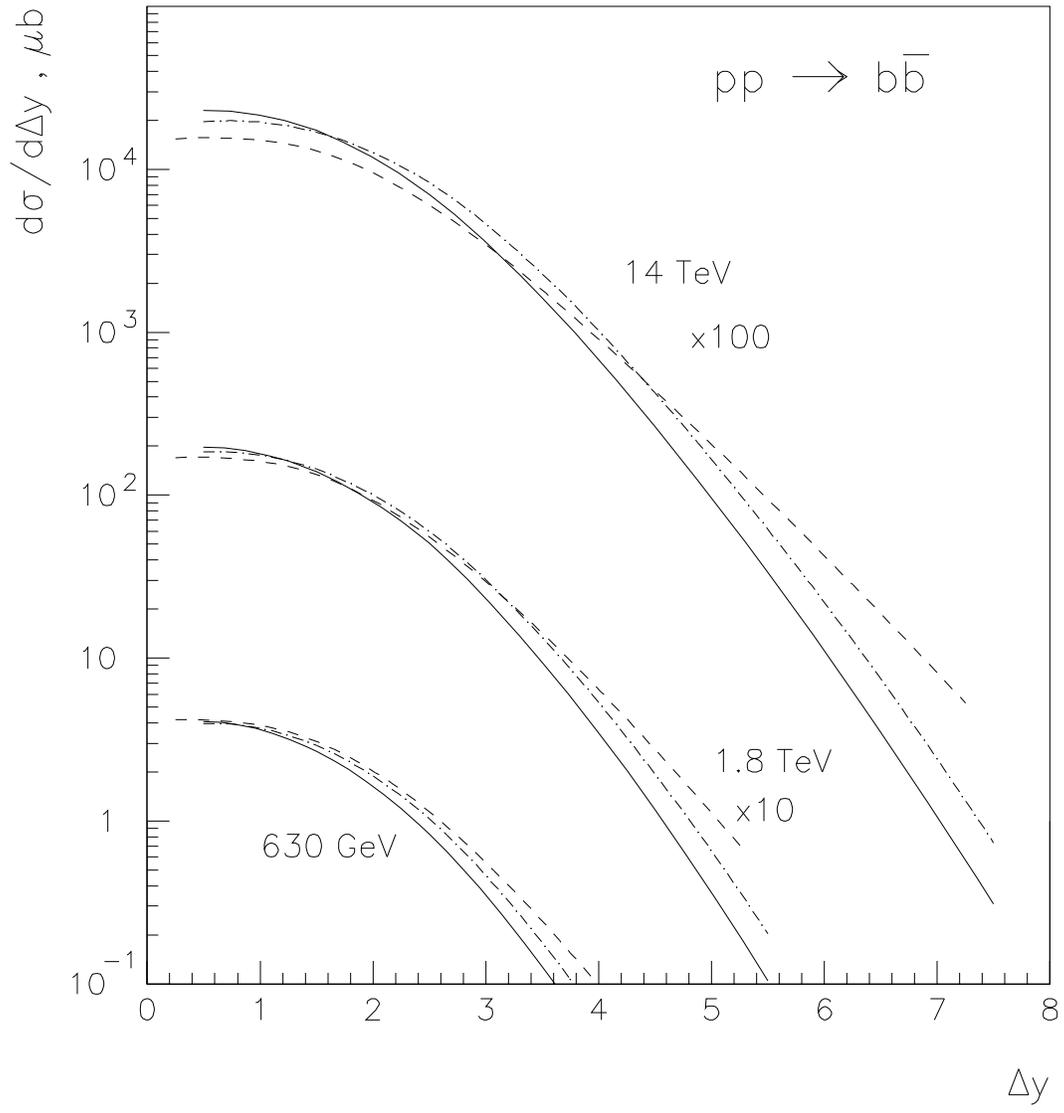
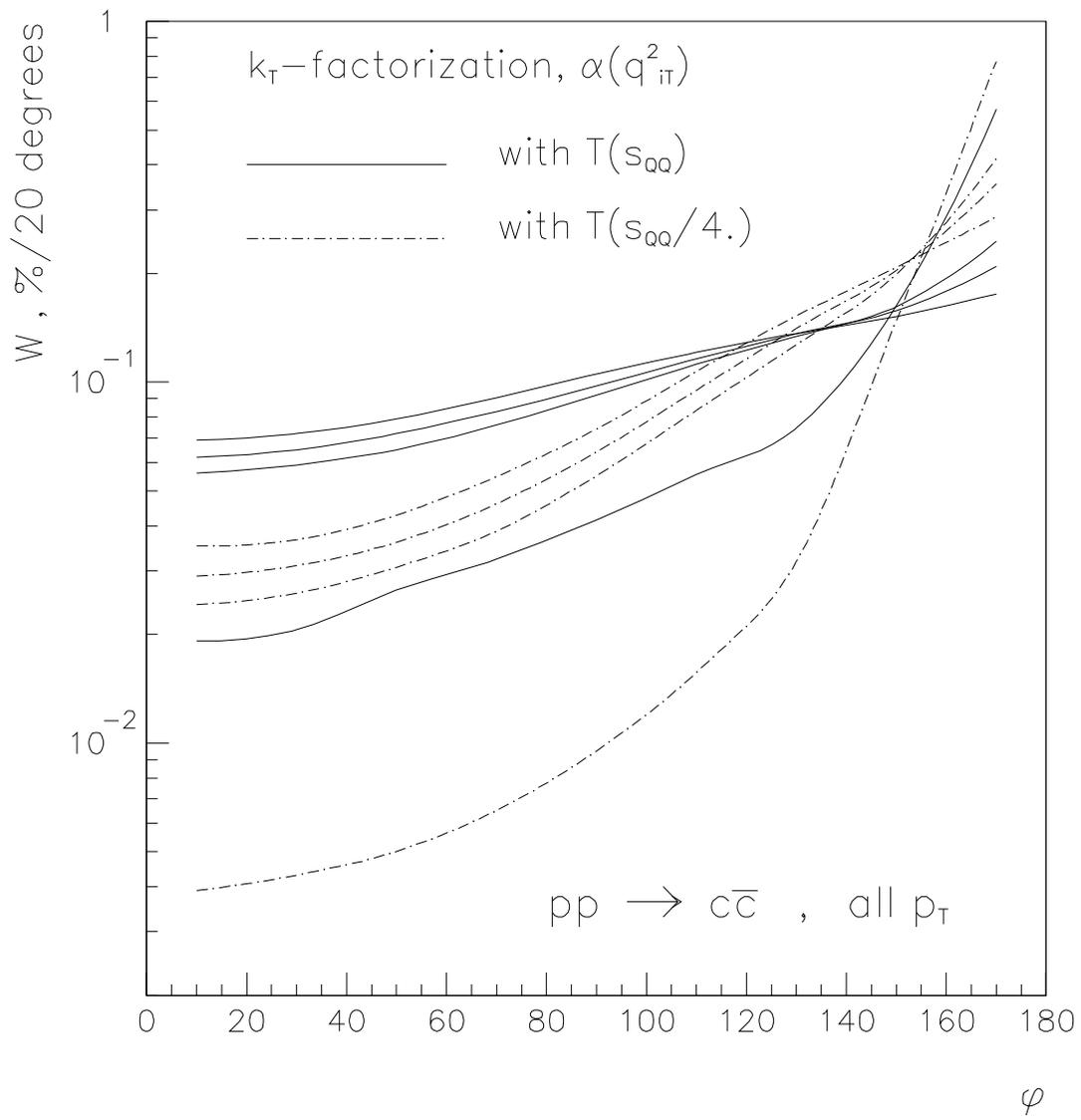
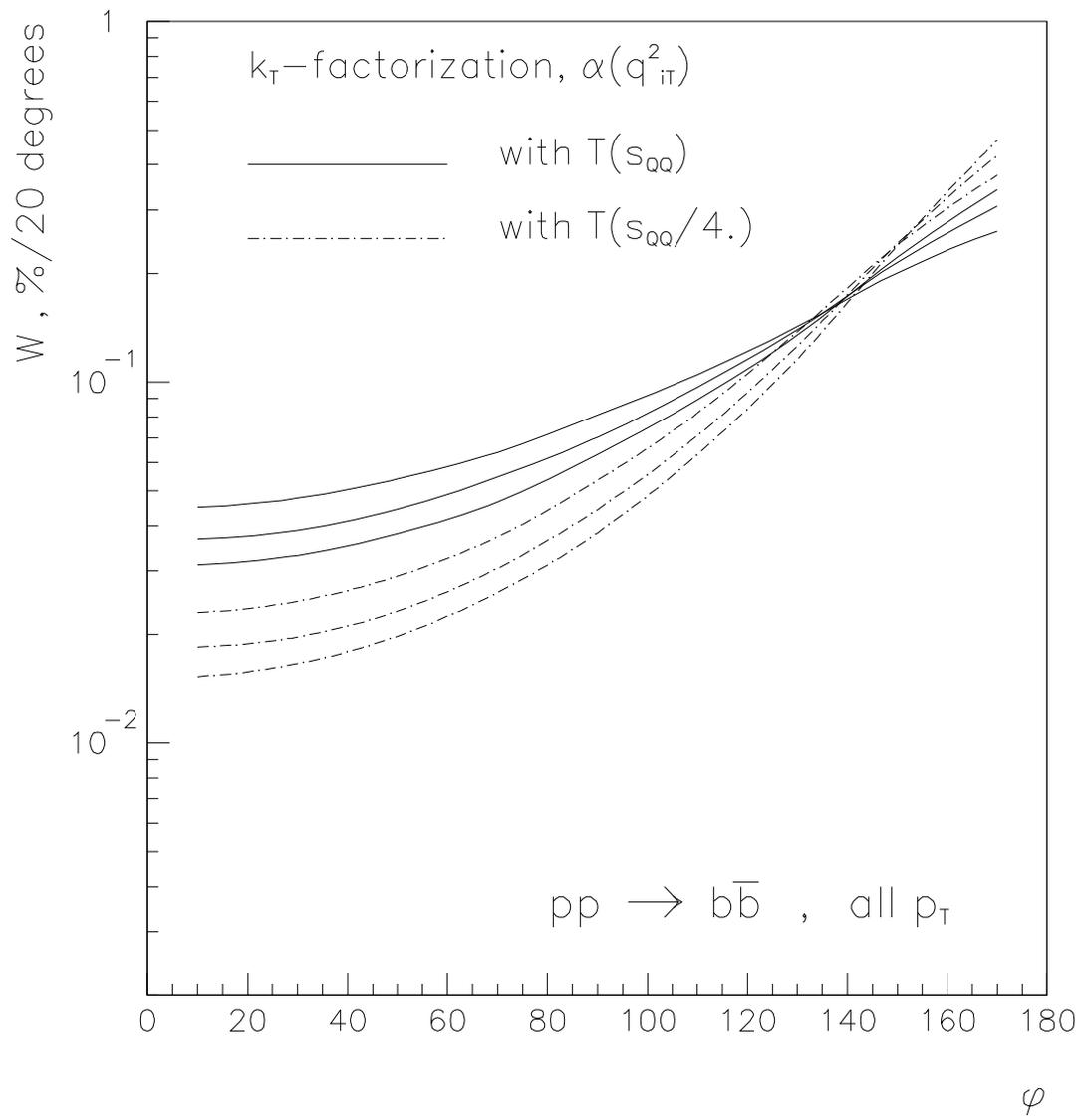
\end{document}